\documentclass[12pt]{article}
\usepackage{geometry}                
\usepackage{graphicx}
\usepackage{bm}
\usepackage{amssymb,amsmath,amsthm}
\usepackage{mathrsfs} 
\def\TODAY{20 January 2011; 5 December 2011}
\title{\vskip-2cm \bf Compound transfer matrices:\\
Constructive and destructive interference}
\author{{\Large Petarpa Boonserm}\\[5pt]
Department of Mathematics and Computer Science\\ 
Faculty of Science, Chulalongkorn University\\
Phayathai Road, Pathumwan\\
Bangkok 10330, Thailand\\[5pt]
{\sf \small petarpa.boonserm@gmail.com} \\[7pt]
{\Large Matt Visser}\\[5pt]
School of Mathematics, Statistics, and Operations Research\\
Victoria University of Wellington \\
PO Box 600, Wellington 6140, New Zealand\\[5pt]
{\sf \small matt.visser@msor.vuw.ac.nz}  }
\date{\TODAY;  \LaTeX-ed  \today}                                           
\begin{document}
\maketitle
\begin{abstract}
{\small
Scattering from a compound barrier, one composed of a number of distinct non-overlapping sub-barriers, has a number of interesting and subtle mathematical features. If one is scattering classical particles, where the wave aspects of the particle can be ignored, the transmission probability of the compound barrier is simply given by the product of the transmission probabilities of the individual sub-barriers.  In contrast if one is scattering waves (whether we are dealing with either purely classical waves or quantum Schrodinger wavefunctions) each sub-barrier contributes phase information (as well as a transmission probability), and these phases can lead to either constructive or destructive interference, with the transmission probability oscillating between nontrivial upper and lower bounds. In this article we shall study these upper and lower bounds in some detail, and also derive bounds on the closely related process of quantum excitation (particle production) via parametric resonance. 
}

\enlargethispage{30pt}
\bigskip

\end{abstract}

\def\d{{\mathrm{d}}}
\newcommand{\scri}{\mathscr{I}}
\newcommand{\sun}{\ensuremath{\odot}}
\def\J{{\mathscr{J}}}
\def\sech{{\mathrm{sech}}}
\clearpage
\tableofcontents
\clearpage

\section{Background}

Consider a general one-dimensional scattering problem. 
One might be interested in classical waves (acoustic waves in a pipe, surface waves in a channel, electromagnetic waves in a waveguide), or quantum waves (the Schrodinger equation).  One formalism that is very well adapted to addressing this mathematical situation is that of ``transfer matrices'' where one relates the waves on the left of a barrier to the waves on the right of the barrier via a $2\times2$ complex transfer matrix. These transfer matrix techniques are discussed, at varying levels of detail, in the textbooks by Merzbacher~\cite{Merzbacher}, Mathews and Venkatesan~\cite{Mathews}, and Singh~\cite{Singh}, and in the pedagogical articles by Sanchez-Soto et al~\cite{Sanchez-Soto} and the present authors~\cite{pedagogical}. 
Other more technical research articles using this formalism include those of Peres~\cite{Peres},  Kowalski and Fry~\cite{Kowalski}, Korasani and Adibi~\cite{Korasani},  and Barriuso et al~\cite{SS1,SS2}.
These techniques have also served as backdrop to some previous results reported by the second author in~\cite{Bounds0}, and related work by both current authors reported in~\cite{Bounds-beta, Bounds-greybody, Bounds-mg, Bounds-analytic,  Shabat-Zakharov, Bounds-thesis, QNF:analytic}.

We start by noting that transfer matrices are of the form:
\begin{equation}
M = \left[ \begin{array}{cc} \alpha & \beta \\ \beta^* & \alpha^* \end{array} \right]; \qquad \qquad  |\alpha|^2 - |\beta|^2 = 1,
\end{equation}
where $\alpha$ and $\beta$ are known as ``Bogoliubov coefficients''. In terms of the perhaps more common transmission and reflection amplitudes one has
\begin{equation}
M  = \left[ \begin{array}{cc} {1/ t} & {r/ t} \\ {r^*/ t^*} & {1/ t^*} \end{array} \right]; \qquad \qquad |t|^2 + |r|^2 = 1.
\end{equation}
The transmission and reflection probabilities are simply
\begin{equation}
T = |t|^2; \qquad\qquad R= |r|^2; \qquad \qquad  T + R = 1.
\end{equation}
Because of these normalization properties we can always find real numbers $\{\theta,\varphi_\alpha, \varphi_\beta\}$ to write
\begin{equation}
\alpha = \cosh(\theta) \exp(i\varphi_\alpha); \qquad \beta= \sinh(\theta) \exp(i\varphi_\beta);
\end{equation}
and
\begin{equation}
t  = \sech(\theta) \exp(-i\varphi_\alpha); \qquad r= \tanh(\theta) \exp(-i[\varphi_\alpha-\varphi_\beta]).
\end{equation}
Now consider a situation where one has a number of non-overlapping barriers. We can assign a distinct transfer matrix $M_i$ to each barrier. How do these transfer matrices combine to give the ``total'' transfer matrix for the compound barrier? 

\clearpage
\noindent
Two key standard results are~\cite{Merzbacher, Mathews, Singh, Sanchez-Soto, pedagogical}:
\begin{itemize}
\item 
If one moves any one of the sub-barriers a distance $a_i$, the effect of this translation is to modify
\begin{equation}
M_i \to  \tilde M_i = \left[ \begin{array}{cc} e^{+i ka_i} & 0 \\ 0 &  e^{-ika_i} \end{array} \right]
M_i
\left[ \begin{array}{cc}  e^{-ika_i} & 0 \\ 0 &  e^{+ika_i} \end{array} \right]
=
\left[ \begin{array}{cc} \alpha_i & \beta_i e^{+2ika_i} \\ \beta_i^* e^{-2ika_i}& \alpha_i^* \end{array} \right].
\end{equation}
That is, the Bogoliubov coefficient $\alpha_i$ (and so the transmission amplitude $t_i$) is invariant under a shift in the physical location of the barrier, while 
 the Bogoliubov coefficient $\beta_i$ (and so the reflection amplitude $r_i$) picks up a shift-dependent phase $e^{+ika_i}$.

\item
A compound transfer matrix (for $n$ distinct non-overlapping localized barriers) is of the form
\begin{equation}
M_{12\dots n} =  M_1 M_2 \dots M_n.
\end{equation}
The order in which these matrices are multiplied together is important.
\end{itemize}
The various phases $e^{i\varphi_{\alpha_i}}$, $e^{i\varphi_{\beta_i}}$, and $e^{+ika_i}$, have a very strong influence on the overall transmission and reflection properties, and can lead to either constructive interference or destructive interference. (In extreme cases one may encounter transmission resonances [reflection-less potentials] where the transmission probability is unity, $T\to 1$, even for rather nontrivial potentials.) 

We shall be interested in extracting as much information as possible regarding the compound transmission probability $T_{12\dots n}$ without having access to knowledge of the individual phases  $e^{i\varphi_{\alpha_i}}$, $e^{i\varphi_{\beta_i}}$, and $e^{+ika_i}$. That is, given only the individual transmission  probabilities $T_i$, what can one say about the total transmission probability  $T_{12\dots n}$? Naturally, information regarding transmission probabilities will also translate into information regarding reflection probabilities. 

We have so far phrased things in terms of a scattering problem, but if one moves from the space domain into the time domain, the same analysis will give information regarding quantum particle production due to parametric resonance.  Each ``barrier'' is now viewed as a time-dependent parametric change in oscillation frequency, and each of these ``parametric excitation episodes'' would by itself excite a number $N_i$ of quanta, where
\begin{equation}
N_i = |\beta_i|^2.
\end{equation}
But, because of the phases $e^{i\varphi_{\alpha_i}}$, $e^{i\varphi_{\beta_i}}$, and $e^{+ika_i}$,  distinct parametric excitation episodes can interfere either constructively or destructively, and the total number of quanta excited,  $N_{12\dots n}$, will be some complicated function of the individual $N_i$ and the phases.  (In extreme cases one may encounter total destructive interference, where the nett (overall) particle production is zero, $N\to 0$, even for a rather nontrivial parametric excitation.) 

We shall also be interested in extracting as much information as possible regarding the nett particle production $N_{12\dots n}$ without having access to knowledge of the individual phases  $e^{i\varphi_{\alpha_i}}$, $e^{i\varphi_{\beta_i}}$, and $e^{+ika_i}$. That is, given only the individual particle production numbers $N_i$, what can one say about the total particle production $N_{12\dots n}$? 

In counterpoint to these phase-dependent questions, if one is working in the limit where the objects being scattered can be treated as classical particles, so their wave properties can be neglected, then the individual transmission probabilities $T_i$ are combined according to the standard rules of ordinary probability theory:
\begin{equation}
T_\mathrm{particles} = \prod_{i=1}^n T_i.
\end{equation}
In the discussion below we will investigate these questions in considerable detail, developing very general upper and lower bounds on the nett Bogoliubov coefficients, nett transmission and reflection probabilities, and nett number of excited quanta in terms of the properties of the individual sub-barriers and/or parametric resonance episodes. 

We have also investigated the possibility of utilizing the Abeles matrix formulation of the scattering problem. (See for instance~\cite{Abeles1, Abeles2, Lekner, Furman}.) While the transfer matrix approach and the Abeles matrix approach are fundamentally just different mathematical ways of addressing the same physical problem, and so must ultimately give completely equivalent answers, the the details of the matrix manipulations involved are different. Specifically, the process of extracting a transmission amplitude from an Abeles matrix  is somewhat messier than that for extracting a transmission amplitude from a transfer matrix --- for this reason we focus on the transfer matrix approach throughout the current article.

\section{Two-barrier systems}

The compound transfer matrix (for two non-overlapping localized barriers) is of the form
\begin{equation}
M_{12} =  M_1 M_2 =  \left[ \begin{array}{cc} \alpha_1 \alpha_2 + \beta_1 \beta_2^*  & \alpha_1 \beta_2 + \beta_1 \alpha_2^* \\
 \alpha_1^* \beta_2^* + \beta_1^* \alpha_2  & \alpha_1^* \alpha_2^* + \beta_1^* \beta_2  \end{array} \right].
\end{equation}
That is:
\begin{equation}
\alpha_{12} =  \alpha_1 \alpha_2 + \beta_1 \beta_2^*; \qquad \beta_{12} =  \alpha_1 \beta_2 + \beta_1 \alpha_2^*.
\end{equation}

\subsection{Bounding the Bogoliubov coefficients}

This leads to the immediate upper bounds
\begin{equation}
|\alpha_{12}| \leq  |\alpha_1| |\alpha_2| + |\beta_1| |\beta_2| ; \qquad |\beta_{12}| \leq  |\alpha_1| |\beta_2| + |\beta_1| |\alpha_2|.
\end{equation}
Now using the normalization conditions, and writing $|\alpha_i| = \cosh\theta_i$ and $|\beta_i| = \sinh\theta_i$,  we have
\begin{equation}
|\alpha_{12}| \leq  \cosh\theta_1 \cosh\theta_2 + \sinh\theta_1\sinh\theta_2 = \cosh(\theta_1+\theta_2), 
\label{E:b1}
\end{equation}
and 
\begin{equation}
|\beta_{12}| \leq  \cosh\theta_1 \sinh\theta_2 + \sinh\theta_1\cosh\theta_2 = \sinh(\theta_1+\theta_2),
\label{E:b2}
\end{equation}
which can be rewritten as
\begin{equation}
|\alpha_{12}| \leq \cosh\{ \cosh^{-1}|\alpha_1| + \cosh^{-1}|\alpha_2|\}; 
\end{equation}
and
\begin{equation}
|\beta_{12}| \leq \sinh\{ \sinh^{-1}|\beta_1| + \sinh^{-1}|\beta_2| \}.
\end{equation}
In the other direction we have the immediate lower bound
\begin{equation}
|\alpha_{12}| \geq  |\alpha_1| |\alpha_2| - |\beta_1| |\beta_2|,
\label{E:b3}
\end{equation}
which we can rewrite as
\begin{equation}
|\alpha_{12}| \geq \cosh\{ \cosh^{-1}|\alpha_1| - \cosh^{-1}|\alpha_2|\}.
\end{equation}
Because cosh is an even function of its argument, there is no need to worry about the relative sizes of the two $\cosh^{-1}|\alpha_i|$ terms.
In contrast, we have to be particularly careful when deriving a lower bound for $\beta_{12}$. (This will be a recurring theme in the article, so we shall be particularly explicit in deriving this result.) Note that
\begin{equation}
\beta_{12} =  \alpha_1 \beta_2 + \beta_1 \alpha_2^* 
= |\alpha_1| |\beta_2| \exp\left(i [\varphi_{\alpha_1}+\varphi_{\beta_2}]) + |\beta_1||\alpha_2| \exp(i [\varphi_{\beta_1}-\varphi_{\alpha_2}]\right).
\end{equation}
But then
\begin{equation}
|\beta_{12}|
=  
\Big| 
|\alpha_1| |\beta_2| + |\beta_1||\alpha_2| \exp(i [\varphi_{\beta_1}-\varphi_{\beta_2}-\varphi_{\alpha_1}-\varphi_{\alpha_2}]) 
\Big |.
\end{equation}
But $|\alpha_1| |\beta_2|$ and $ |\beta_1||\alpha_2|$ are both real and positive --- so the quantity above is minimized when destructive interference is maximum. It is now easy to see that this occurs when the phases are such that 
\begin{equation}
\varphi_{\beta_1}-\varphi_{\beta_2}-\varphi_{\alpha_1}-\varphi_{\alpha_2} =  (2n+1) \pi,
\end{equation}
in which case
\begin{eqnarray}
 \exp(i [\varphi_{\beta_1}-\varphi_{\beta_2}-\varphi_{\alpha_1}-\varphi_{\alpha_2}])  = - 1.
\end{eqnarray}
Consequently 
\begin{equation}
|\beta_{12}| \geq  \Big| \; |\alpha_1| |\beta_2| - |\beta_1| |\alpha_2| \; \Big|.
\end{equation}
Note that the outermost set of absolute value signs is particularly critical, guaranteeing that the RHS is non-negative.
This can now be rewritten as
\begin{equation}
|\beta_{12}| \geq \sinh\Big| \; \sinh^{-1}|\beta_1| - \sinh^{-1}|\beta_2| \; \Big|.
\label{E:b4}
\end{equation}
Again, the outermost set of absolute value signs is particularly critical, guaranteeing that the RHS is non-negative.

\subsection{Bounding transmission and reflection probabilities}
\label{S:2-2}

Now working in terms of transmission and reflection amplitudes and probabilities we have
\begin{equation}
M = \left[ \begin{array}{cc} \alpha & \beta \\ \beta^* & \alpha^* \end{array} \right] = 
 \left[ \begin{array}{cc} {1/ t} & {r/ t} \\ {r^*/ t^*} & {1/ t^*} \end{array} \right],
\end{equation}
and
\begin{equation}
T =|t|^2 = {1\over |\alpha|^2}= \sech^2\theta; \qquad  R = |r|^2 = {|\beta|^2\over|\alpha|^2} =\tanh^2\theta.
\end{equation}
The upper bounds on the Bogoliubov coefficients lead to a lower bound on $T$ and an upper bound on $R$ as follows: 
\begin{equation}
T_{12} \geq \sech^2\left\{ \sech^{-1}\sqrt{T_1} +  \sech^{-1}\sqrt{T_2} \right\}; 
\end{equation}
\begin{equation}
R_{12} \leq \tanh^2\left\{ \tanh^{-1}\sqrt{R_1} +  \tanh^{-1}\sqrt{R_2} \right\}.
\end{equation}
Similarly, the lower bounds on the Bogoliubov coefficients lead to an upper bound on $T$ and a lower bound on $R$ as follows: 
\begin{equation}
T_{12} \leq \sech^2\left\{ \sech^{-1}\sqrt{T_1} -  \sech^{-1}\sqrt{T_2} \right\}; 
\end{equation}
\begin{equation}
R_{12} \geq \tanh^2\left\{ \tanh^{-1}\sqrt{R_1} -  \tanh^{-1}\sqrt{R_2} \right\}.
\end{equation}
Note that, because we are squaring the sech and tanh,  we do not need to worry about the relative magnitudes of the $T_i$ and the $R_i$ in the two formulae above.

By manipulating the hyperbolic and inverse hyperbolic functions, (see some key hyperbolic identities in appendix~\ref{A:hyperbolic}), these bounds can be brought into the rational algebraic form
\begin{equation}
T_{12} \geq {T_1 T_2\over  \left\{ 1 + \sqrt{1-T_1}\sqrt{1-T_2} \right\}^2}; 
\qquad 
R_{12} \leq \left\{ {\sqrt{R_1} + \sqrt{R_2} \over 1 + \sqrt{R_1}\sqrt{R_2} }\right\}^2.
\end{equation}
Similarly
\begin{equation}
T_{12} \leq {T_1 T_2\over  \left\{ 1 - \sqrt{1-T_1}\sqrt{1-T_2} \right\}^2}; 
\qquad 
R_{12} \geq \left\{ {\sqrt{R_1} - \sqrt{R_2} \over 1 - \sqrt{R_1}\sqrt{R_2} }\right\}^2.
\end{equation}
As a useful internal consistency check on the formalism note that for $R_i\in[0,1]$ we have
\begin{equation}
 {\sqrt{R_1} + \sqrt{R_2} \over 1 + \sqrt{R_1}\sqrt{R_2} } \leq 1,
\end{equation}
and that for $T_i\in [0,1]$ we have
\begin{equation}
 {T_1 T_2\over  \left\{ 1 - \sqrt{1-T_1}\sqrt{1-T_2} \right\}^2} \leq 1.
\end{equation}
Note that to get an exact transmission resonance $T_{12}=1$ it is necessary (but certainly not sufficient) that 
\begin{equation}
 {T_1 T_2\over  \left\{ 1 - \sqrt{1-T_1}\sqrt{1-T_2} \right\}^2} = 1.
\end{equation}
But this implies $T_1=T_2$. That is, if one ever wishes to obtain an exact transmission resonance from two disjoint non-overlapping barriers, then the two individual barriers must have equal transmission probabilities.

\subsection{Bounding particle production}
\label{S:2-3}

If in contrast we work in a particle production scenario, (via episodic parametric resonance), where $N=|\beta|^2$ we can similarly extract upper and lower bounds
\begin{equation}
N_{12} \leq \sinh^2\left\{  \sinh^{-1} \sqrt{N_1}+   \sinh^{-1}\sqrt{N_2}\right\};
\end{equation}
\begin{equation}
N_{12} \geq \sinh^2\left\{ \sinh^{-1} \sqrt{N_1}-   \sinh^{-1}\sqrt{N_2} \right\}.
\end{equation}
Again, because we are squaring the sinh,  we do not need to worry about the relative magnitudes of the $N_i$ in the formula above.
These bounds can be converted (see appendix~\ref{A:hyperbolic}) to the algebraic statements
\begin{equation}
N_{12} \leq \left\{\sqrt{N_1(N_2+1)}+   \sqrt{N_2(N_1+1)}\right\}^2;
\end{equation}
\begin{equation}
N_{12} \geq \left\{\sqrt{N_1(N_2+1)}-   \sqrt{N_2(N_1+1)}\right\}^2.
\end{equation}
Note that if one wishes to get complete destructive interference, then one must have $\left\{\sqrt{N_1(N_2+1)}-   \sqrt{N_2(N_1+1)}\right\} = 0$, implying $N_1=N_2$.  That is, if one ever wishes to obtain an exact cancellation of particle production from two disjoint non-overlapping parametric resonance episodes, then the two individual episodes must individually have equal quantities of particle production.

\subsection{Summary}

Having worked through the two barrier case in some detail, we are now in a position to consider the more general case. Note that all the bounds above have been carefully checked against the extant literature, (see for example~\cite{Merzbacher, Mathews, Singh, Sanchez-Soto, pedagogical, Peres, Kowalski, Korasani, SS1, SS2, Bounds0, Bounds-beta, Bounds-greybody, Bounds-mg, Bounds-analytic,  Shabat-Zakharov, Bounds-thesis, QNF:analytic}), and are compatible with all known results. 
For the multiple-barrier situation, many of the calculations are immediate and straightforward generalizations of the above --- we shall soon see that the tricky one is the lower bound on the (absolute magnitude of the) Bogoliubov coefficients $|\alpha_{12\dots n}|$ and $|\beta_{12\dots n}|$, (which ultimately leads to the upper bound on $T_{12\dots n}$ and the lower bound on $N_{12\dots n}$).

\section{General upper bounds on $|\alpha|$ and $|\beta|$}

The two-barrier upper bounds on  the Bogoliubov coefficients $|\alpha|$ and $|\beta|$ immediately generalize (by straightforward iteration)  to the composition of $n$ transfer matrices 
\begin{equation}
|\alpha_{12\dots n}| \leq \cosh\left\{ \sum_{i=1}^n \cosh^{-1}|\alpha_i| \right\}; 
\qquad 
|\beta_{12\dots n}| \leq \sinh\left\{ \sum_{i=1}^n \sinh^{-1}|\beta_i|  \right\}.
\end{equation}
This immediately leads to a lower bound on transmission probability and an upper bound on reflection probability.
\begin{equation}
T_{12\dots n} \geq \sech^2\left\{ \sum_{i=1}^n \sech^{-1}\sqrt{T_i} \right\}; 
\qquad 
R_{12\dots n} \leq \tanh^2\left\{ \sum_{i=1}^n \tanh^{-1}\sqrt{R_i} \right\}.
\end{equation}
If we work in a particle production scenario where $N=|\beta|^2$ we can similarly write
\begin{equation}
N_{12\dots n} \leq \sinh^2\left\{ \sum_{i=1}^n \sinh^{-1}\sqrt{N_i}  \right\}.
\end{equation}
In the general case these appear to be the most tractable form of the bounds.
(We have also verified the correctness of these results by explicitly iterating the two-barrier results to investigate three-barrier, four-barrier, and certain particularly tractable $n$-barrier systems, comparing with the general result presented above.)
 
\section{General lower bounds on $|\alpha|$ and $|\beta|$}

When it comes to bounding the Bogoliubov coefficients from below we have already seen
\begin{equation}
|\alpha_{12}| \geq  |\alpha_1| |\alpha_2| - |\beta_1| |\beta_2| ; \qquad |\beta_{12}| \geq  \Big|\;|\alpha_1| |\beta_2| - |\beta_1| |\alpha_2|\; \Big|; 
\end{equation}
which can be rewritten as
\begin{equation}
|\alpha_{12}| \geq \cosh\{ \cosh^{-1}|\alpha_1| - \cosh^{-1}|\alpha_2|\}; 
\qquad 
|\beta_{12}| \geq \sinh\Big| \;\sinh^{-1}|\beta_1| - \sinh^{-1}|\beta_2| \;\Big|.
\end{equation}
However generalizing these lower bound inequalities to $n$ steps is much more difficult. We shall build up our lower bound in stages, first providing a recursive version of the bound before constructing an explicit solution to the recursion relation.

\subsection{Implicit iterative lower bounds on $|\alpha|$ and $|\beta|$}

Define a collection of $n$ parameters:
\begin{equation}
\theta_i = \cosh^{-1}|\alpha_i|,
\end{equation}
and the sums ($m\in\{1,2,3,\dots n\}$)
\begin{equation}
S_m = \sum_{i=1}^m \theta_i.
\end{equation}
Now take
\begin{equation}
B_1 = \theta_1,
\end{equation}
and, for $m\in\{1,2,3,\dots n-1\}$, iteratively define the quantities $B_{m+1}$ by
\begin{equation}
B_{m+1} =  (\theta_{m+1} - S_m) \; H(\theta_{m+1} - S_m) + (B_m-\theta_{m+1}) \; H(B_m-\theta_{m+1}),
\end{equation}
where $H(\cdot)$ is the Heaviside step function. 

\paragraph{Theorem:} By iterating the 2-step bounds one has
\begin{equation}
|\alpha_{12\dots n}| \geq \cosh B_n; \qquad |\beta_{12\dots n}| \geq \sinh B_n; 
\end{equation}
and consequently
\begin{equation}
T_{12\dots n} \leq \sech^2 B_n; \qquad R_{12\dots n} \geq \tanh^2 B_n;  \qquad N_{12\dots n}  \geq \sinh^2 B_n.
\end{equation}

\paragraph{Proof by induction:} 
When we iterate the definition for $B_n$ the first two times we obtain
\begin{eqnarray}
S_1 &=& \theta_1; \qquad \qquad B_1 = \theta_1;
\\
S_2 &=& \theta_1+\theta_2; \qquad B_2 = |\theta_1-\theta_2|.
\end{eqnarray}
Thus by our previous results, the claimed theorem is certainly true for $n=2$. Now apply mathematical induction: Assume 
that at each stage the interval $[B_m,S_m]$ characterizes the  highest possible and lowest possible values of $\theta_{12\dots m}$.  Applying the 2-step bound to the pair $\theta_{12\dots m}$ and $\theta_{m+1}$ leads trivially to $\theta_{12\dots(m+1)}$ being bounded from above by
\begin{equation}
S_{m+1} = S_m + \theta_{m+1},
\end{equation}
and less trivially to being bounded from below by
\begin{equation}
B_{m+1} =  (\theta_{m+1} - S_m) \; H(\theta_{m+1} - S_m) + (B_m-\theta_{m+1}) \; H(B_m-\theta_{m+1}),
\end{equation}
as asserted above. This completes the inductive step. That is:
\begin{equation}
\theta_{12\dots(m+1)} \in [B_{m+1}, S_{m+1}].
\end{equation}
\hfill $\Box$\\
However these bounds are unfortunately defined in a relatively messy and implicit iterative manner --- can this be usefully simplified? Can we make the bounds explicit?

\subsection{Symmetry properties for the lower bound}

When we iterate the general definition of $B_n$,  previously used to obtain $B_2$, one more time we see
\begin{eqnarray*}
S_3 &=& \theta_1+\theta_2+\theta_3; \qquad B_3 =
\max\{ \theta_3 - (\theta_1+\theta_2),  |\theta_1-\theta_2|-\theta_3, 0\}.
\end{eqnarray*}
Is there any further way of simplifying this? Rewrite $B_3$ as
\begin{equation}
B_3 = \max\{  \theta_1-\theta_2-\theta_3, \;\theta_2-\theta_3-\theta_1, \; \theta_3 - \theta_1-\theta_2, \;0\}.
\end{equation}
Note that this form of $B_3$ is manifestly symmetric under arbitrary permutations of the labels $123$. 
One suspects that there is a good reason for this.  In fact there is.

\paragraph{Theorem:}  $\forall n$ $ B_n(\theta_i)$ is a totally symmetric function of the $n$ parameters $\theta_i$. 

\paragraph{Proof:} Note that the individual transfer matrices $M_i$ and $M^T_i$ trivially have the same values of $S_i$ and $B_i$, and in fact have the same values of $\theta_i$. Note further that for any two transfer matrices $M_1 M_2$ and $M_2 M_1$ have identical values of $S_2$ and $B_2$.  Finally for any two transfer matrices the products $M_1 M_2^T$ and $M_1 M_2$ have identical values of $S_2$ and $B_2$. (These assertions all follow from the simple results obtained above for compounding two transfer matrices.)

That is:
\begin{equation}
S(M_i) = S(M_i^T); \qquad B(M_i) = B(M^T_i);
\end{equation}
\begin{equation}
S(M_1 M_2) = S(M_2 M_1); \qquad B(M_1 M_2) = B(M_2 M_1);
\end{equation}
\begin{equation}
S(M_1 M_2^T) = S(M_1 M_2); \qquad B(M_1 M_2^T) = B(M_1 M_2).
\end{equation}

But then by cyclic permutation
\begin{equation}
B(M_1 M_2 M_3) = B(M_3 M_1 M_2) = B(M_2 M_3 M_1),
\end{equation}
and furthermore
\begin{eqnarray}
B(M_2 M_1 M_3) &=& B((M_1^T M_2^T)^T M_3) 
\nonumber\\
&=&
 B((M_1^T M_2^T) M_3) 
 \nonumber\\
&=&
 B(M_1^T (M_2^T M_3))  
\nonumber\\
&=&
 B(M_1 ( M_2^T M_3)) 
\nonumber\\
&=&
 B( (M_2^T M_3) M_1) 
\nonumber\\
&=&
 B(M_2^T (M_3 M_1))  
\nonumber\\
&=&
B(M_2 M_3 M_1) 
\nonumber\\
&=&B(M_1 M_2 M_3).
\end{eqnarray}
That is
\begin{equation}
B(M_2 M_1 M_3)  = B(M_1 M_2 M_3).
\end{equation}
Combining these results implies that $B(M_1M_2M_3)$ is a symmetric function of the three transfer matrices $M_i$, and hence a symmetric function of the three parameters $\theta_1$, $\theta_2$, $\theta_3$.
But this argument now generalizes --- For any $B(M_1M_2\dots M_n)$ you can use this argument to show
\begin{equation}
B(M_1M_2 M_3 \dots M_n) = B(M_2M_1 M_3 \dots M_n); 
\end{equation}
and
\begin{equation}
 B(M_1M_2 M_3 \dots M_n) = B(M_2 M_3 \dots M_n M_1),
\end{equation}
which implies complete symmetry in all its $n$ arguments $\theta_i$. \hfill $\Box$\\
Based on this observation we can now assert a bolder theorem that has the effect of yielding an explicit (non-iterative) formula for $B_n$.

\subsection{Explicit non-iterative formula for the lower bound}

\paragraph{Theorem:}
\begin{equation}
\forall m : B_{m} =  \max_{i\in\{1,2,\dots m\}}\{ 2 \theta_i - S_m, 0\}.
\end{equation}
\paragraph{Proof by induction:}
We have already seen that the iterative definition of $B_n$ can be written as
\begin{equation}
B_{m+1} = \max\{\theta_{m+1}-S_m, B_m-\theta_{m+1}, 0\},
\end{equation}
which we can also rewrite as
\begin{equation}
B_{m+1} = \max\{2 \theta_{m+1}-S_{m+1}, B_m-\theta_{m+1}, 0\}.
\end{equation}
Now apply induction. The assertion of the theorem is certainly true for $m=1$ and $m=2$, and has even been explicitly verified for $m=3$. Now assume it holds up to some $m$, then
\begin{eqnarray*}
B_{m+1} &=& \max\{2 \theta_{m+1}-S_{m+1}, B_m-\theta_{m+1}, 0\} 
\\
 &=& \max\left\{2 \theta_{m+1}-S_{m+1}, \max_{i\in\{1,2,\dots m\}}\{ 2 \theta_i - S_m, 0\}  -\theta_{m+1}, 0\right\} 
\\
 &=& \max\left\{2 \theta_{m+1}-S_{m+1}, \max_{i\in\{1,2,\dots m\}}\{ 2 \theta_i - S_{m+1}, 0\}  , 0\right\} 
\\
 &=&\max_{i\in\{1,2,\dots m,(m+1)\}}\{ 2 \theta_i - S_{m+1} , 0\}.
\end{eqnarray*}
This proves the inductive step. Consequently
\begin{equation}
\forall m: B_{m} =  \max_{i\in\{1,2,\dots m\}}\{ 2 \theta_i - S_m, 0\},
\end{equation}
and in particular
\begin{equation}
B_{n} =  \max_{i\in\{1,2,\dots n\}}\{ 2 \theta_i - S_n, 0\}. 
\end{equation}
(For completeness, note that we have explicitly checked the equivalence of these  iterative and non-iterative results for $B_n$ by using symbolic manipulation packages for multiple examples for the cases $n=5$ and $n=10$.)  \hfill $\Box$

To simplify the formalism even further, let us now define
\begin{equation}
\theta_\mathrm{peak} = \max_{i\in\{1,2,\dots n\}} \theta_i,
\end{equation}
and
\begin{equation}
\theta_\mathrm{off\,peak} = \sum_{i\neq i_\mathrm{peak}} \theta_i =   \sum_{i=1}^n \theta_i - \theta_\mathrm{peak}= S_n - \theta_\mathrm{peak}.
\end{equation}
(We wish to use the subscript ``peak'' for the maximum of the individual $\theta_i$'s;  the subscripts ``max'' and ``min'' will be reserved for bounds on the $n$-fold composite transfer matrix.)
Then we can write
\begin{equation}
B_{n} =  \max\{ 2\theta_\mathrm{peak} - S_n, 0\},
\end{equation}
or
\begin{equation}
B_{n} =  \max\{ \theta_\mathrm{peak} - \theta_\mathrm{off\,peak}, 0\}.
\end{equation}
Note that the max function is still needed to guarantee that the $B_n$ are non-negative.  With this explicit formula for $B_n$ in hand, we have
\begin{equation}
|\alpha_{12\dots n}| \geq  \cosh\left[ \max\{ 2\theta_\mathrm{peak} - S_n, 0\}\right]; 
\qquad
|\beta_{12\dots n}| \geq  \sinh\left[ \max\{ 2\theta_\mathrm{peak} - S_n, 0\}\right].
\end{equation}

\subsection{Geometrical interpretation}

These lower bounds on the Bogoliubov coefficients can also be given a clean and intuitive geometrical interpretation. From the work of Barriuso,  Monzon, Sanchez-Soto, and Cari\~nena~\cite{Sanchez-Soto, SS1,SS2} it is known that the composition of two barriers can be understood as the composition of two hyperbolic vectors
(translations in the hyperbolic plane,  ``turns'' in Hamilton's original notation). The length of these hyperbolic vectors is related to the Bogoliubov, transmission, and reflection coefficients by
\begin{equation}
\ell_i = \cosh^{-1}|\alpha_i| = \sinh^{-1}|\beta_i| = \sech^{-1}\sqrt{T_i} = \tanh^{-1}\sqrt{R_i}. 
\end{equation}
The two-barrier upper bounds (\ref{E:b1}) and (\ref{E:b2}) can then
be easily derived from geometrical considerations as hyperbolic space applications of the triangle inequalities.
The lower bounds (\ref{E:b3}) and (\ref{E:b4}) are a little more subtle, but for the two-barrier case  both upper and lower bounds can be summarized as
\begin{equation}
|\ell_1 - \ell_2| \leq \ell_{12} \leq \ell_1 + \ell_2.
\end{equation}
For the $n$-barrier case the upper bound is straightforward
\begin{equation}
\ell_{12\dots n} \leq \sum_{i=1}^n \ell_i.
\end{equation}
The lower bound is a little trickier. Separate out the lengths $\ell_i$ of the hyperbolic vectors into the ``largest'' and the ``rest''  (corresponding to what we previously called ``peak'' and ``off peak'').  Then geometrically
\begin{equation}
\ell_{12\dots n} \geq \max\left\{ \ell_\mathrm{largest} - \sum_\mathrm{rest} \ell_i , \; 0\right\}.
\end{equation}
It is only when the single largest step exceeds the maximum possible size of all the other remaining steps put together that one obtains a non-trivial lower bound. 
We shall now apply this formalism to bounding the the transmission and reflection probabilities, and to bounding the amount of particle production. 

\section{Application to $T$, $R$, and $N$}

First, we note that
\begin{equation}
T_{12\dots n} \leq \sech^2 \left\{    \max\{  \theta_\mathrm{peak} -  \theta_\mathrm{off\,peak}, 0\}       \right\},
\end{equation}
or equivalently
\begin{equation}
T_{12\dots n} \leq \sech^2 \left\{   \max\{  2 \theta_\mathrm{peak} - S_n, 0\}       \right\}.
\end{equation}
That is
\begin{equation}
T_{12\dots n} \leq \sech^2 \left\{    \max\left\{ 2 \; \sech^{-1}\sqrt{T_\mathrm{peak}} - \sech^{-1}\sqrt{T_\mathrm{min}}, 0\right\}       \right\} ,
\end{equation}
where we have defined $T_\mathrm{min}$ by 
\begin{equation}
T_{12\dots n} \geq T_\mathrm{min} \equiv  \sech^2 \left\{    S_n    \right\}.
\end{equation}
Second, we note that
\begin{equation}
R_{12\dots n}  \geq \tanh^2 \left\{    \max\{ 2 \theta_\mathrm{peak} - S_n, 0\}       \right\} ,
\end{equation}
that is
\begin{equation}
R_{12\dots n} \geq \tanh^2 \left\{    \max\left\{ 2 \tanh^{-1}\sqrt{R_\mathrm{peak}} - \tanh^{-1}\sqrt{R_\mathrm{max}}, 0\right\}       \right\}, 
\end{equation}
where we have defined $R_\mathrm{max}$ by
\begin{equation}
R_{12\dots n} \leq R_\mathrm{max} \equiv  \tanh^2 \left\{    S_n    \right\}.
\end{equation}
Finally,
\begin{equation}
N_{12\dots n}  \geq \sinh^2 \left\{    \max\{ 2 \theta_\mathrm{peak} - S_n, 0\}       \right\} ,
\end{equation}
that is
\begin{equation}
N_{12\dots n} \geq \sinh^2 \left\{    \max\left\{ 2 \sinh^{-1}\sqrt{N_\mathrm{peak}} - \sinh^{-1}\sqrt{N_\mathrm{max}}, 0\right\}       \right\} ,
\end{equation}
where we have defined $N_\mathrm{max}$ by 
\begin{equation}
N_{12\dots n} \leq N_\mathrm{max} \equiv  \sinh^2 \left\{    S_n    \right\}.
\end{equation}
If we want to be a little more explicit, then we could write these bounds as
\begin{equation}
T_{12\dots n} \leq \sech^2 \left\{    \max\left\{ 2 \; \sech^{-1}\sqrt{T_\mathrm{peak}} - \sum_{i=1}^n\sech^{-1}\sqrt{T_i}, 0\right\}       \right\};
\end{equation}
\begin{equation}
R_{12\dots n} \geq \tanh^2 \left\{    \max\left\{ 2 \tanh^{-1}\sqrt{R_\mathrm{peak}} - \sum_{i=1}^n \tanh^{-1}\sqrt{R_i}, 0\right\}       \right\} ;
\end{equation}
\begin{equation}
N_{12\dots n} \geq \sinh^2 \left\{    \max\left\{ 2 \sinh^{-1}\sqrt{N_\mathrm{peak}} - \sum_{i=1}^n \sinh^{-1}\sqrt{N_i}, 0\right\}       \right\}. 
\end{equation}
This appears to be the simplest form achievable for these three bounds.
Remember that these three bounds, coming from lower bounds on the Bogoliubov coefficients are complemented by three other simpler bounds
\begin{equation}
T_{12\dots n} \geq \sech^2\left\{ \sum_{i=1}^n \sech^{-1}\sqrt{T_i} \right\}; 
\qquad 
R_{12\dots n} \leq \tanh^2\left\{ \sum_{i=1}^n \tanh^{-1}\sqrt{R_i} \right\}; 
\end{equation}
\begin{equation}
N_{12\dots n} \leq \sinh^2\left\{ \sum_{i=1}^n \sinh^{-1}\sqrt{N_i}  \right\}.
\end{equation}
coming from upper bounds on the Bogoliubov coefficients.

\section{Perfect transmission resonances}

Transmission resonances (perfect transmission resonances) occur when $T\to 1$ at certain energies or barrier spacings when the phases work out just right to cancel the reflection. But the  occurrence of these resonances is still constrained by our general bound
\begin{equation}
T_{12\dots n} \leq \sech^2 \left\{    \max\left\{ 2 \sech^{-1}\sqrt{T_\mathrm{peak}} - \sum_{i=1}^n\sech^{-1}\sqrt{T_i}, 0\right\}       \right\}.
\end{equation}
So a perfect transmission resonance can only occur if
\begin{equation}
2 \sech^{-1}\sqrt{T_\mathrm{peak}} - \sum_{i=1}^n\sech^{-1}\sqrt{T_i} \leq 0,
\end{equation}
(this is a necessary condition, not a sufficient condition). This is equivalent to requiring
\begin{equation}
T_\mathrm{peak}
\leq \sech^2\left\{  {1\over2} \sum_{i=1}^n\sech^{-1}\sqrt{T_i} \right\}\equiv 
\sech^2\left\{  {1\over2} \sech^{-1}\sqrt{T_\mathrm{min}} \right\}.
\end{equation}
Now use the ``half angle formula'' 
\begin{equation}
\sech(x/2) = \sqrt{2\over\cosh x + 1} = \sqrt{ 2\; \sech x \over 1 + \sech x},
\end{equation}
to obtain 
\begin{equation}
T_\mathrm{peak}
\leq  {2\sqrt{T_\mathrm{min}} \over (1 +  \sqrt{T_\mathrm{min}} ) }.
\end{equation}
Again, this is a necessary condition (not a sufficient condition) in order for a perfect transmission resonance to be possible.

In terms of a particle production scenario, a transmission resonance translates to ``complete destructive interference'' between two or more particle creation events (so that the nett particle production is zero). To be absolutely certain of avoiding  ``complete destructive interference'' one must have
\begin{equation}
N_\mathrm{peak}
> \sinh^2\left\{  {1\over2} \sum_{i=1}^n\sinh^{-1}\sqrt{N_i} \right\}. 
\end{equation}
But we know
\begin{equation}
N \leq N_\mathrm{max} \equiv \sinh^2 S_n =  \sinh^2\left\{  \sum_{i=1}^n\sinh^{-1}\sqrt{N_i} \right\}, 
\end{equation}
so we can rewrite this as
\begin{equation}
N_\mathrm{peak}
>  {\sqrt{N_\mathrm{max}+1}-1\over2}.
\end{equation}
This is a sufficient condition for $N_\mathrm{min} >0$, a sufficient condition for there to be at least some overall particle production.

\section{Discussion}

We have seen how a rather nontrivial application of the transfer matrix formalism, (see for example~\cite{Merzbacher, Mathews, Singh, Sanchez-Soto, pedagogical} for basic background information), allows us to place rather nontrivial bounds on the physics of compound scattering systems --- even in the absence of detailed information concerning their internal structure.
The compound scattering systems we are interested in are one-dimensional systems built up from a number of disjoint non-overlapping barriers; for such systems we have explicitly shown that the nett scattering properties (unsurprisingly) depend not only on the transmission probability, $T_i$, of each individual barrier, but also on a number of phases associated with these individual barriers and the separation between them.  Surprisingly, even in the absence of quantitative information regarding the value of these phases it is nevertheless possible to place rigorous and nontrivial upper and lower bounds on the scattering properties of the compound system.  The resulting bounds also apply (with suitable modification of the language) to bounds on the number of quanta excited via parametric resonance.  These bounds have all been carefully checked against the extant literature, (see for example~\cite{Merzbacher, Mathews, Singh, Sanchez-Soto, pedagogical, Peres, Kowalski, Korasani, SS1, SS2, Bounds0, Bounds-beta, Bounds-greybody, Bounds-mg, Bounds-analytic,  Shabat-Zakharov, Bounds-thesis, QNF:analytic}), and are compatible with all known results. 
The bounds are explicit, compact, and (despite their relative simplicity) appear to be entirely novel.

\section*{Acknowledgments}
This research was supported by the Marsden Fund, administered by the Royal Society of New Zealand. PB was additionally supported by a scholarship from the Royal Government of Thailand, and partially supported by a travel grant from FQXi, and by a grant for the professional development of new academic staff from the Ratchadapisek Somphot Fund at Chulalongkorn University,  and by the Research Strategic plan program (A1B1), Faculty
of Science, Chulalongkorn University. The authors wish to thank the referee for several useful suggestions regarding the presentation and interpretation of the results.

\appendix
\section{Hyperbolic identities}
\label{A:hyperbolic}

Here are several useful and quite elementary hyperbolic identities (particularly relevant to sections~\ref{S:2-2} and \ref{S:2-3}):
\begin{eqnarray}
\sinh\left(\sinh^{-1}A+\sinh^{-1}B\right) &=&  \sinh\left(\sinh^{-1}A\right)  \cosh\left(\sinh^{-1}B\right)
\nonumber\\
&&\qquad  + \cosh\left(\sinh^{-1}A\right)  \sinh\left(\sinh^{-1}B\right)  
\nonumber\\
&=& A \;\sqrt{1+B^2} + \sqrt{1+A^2} \; B.
\end{eqnarray}
\begin{eqnarray}
\cosh\left(\sinh^{-1}A+\sinh^{-1}B\right) &=&  \cosh\left(\sinh^{-1}A\right)  \cosh\left(\sinh^{-1}B\right)
\nonumber\\
&&\qquad  + \sinh\left(\sinh^{-1}A\right)  \sinh\left(\sinh^{-1}B\right)  
\nonumber\\
&=& \sqrt{1+A^2} \;\sqrt{1+B^2} + A \; B.
\end{eqnarray}
\begin{eqnarray}
\cosh\left(\cosh^{-1}A+\cosh^{-1}B\right) &=&  \cosh\left(\cosh^{-1}A\right)  \cosh\left(\cosh^{-1}B\right)
\nonumber\\
&&\qquad  + \sinh\left(\cosh^{-1}A\right)  \sinh\left(\cosh^{-1}B\right)  
\nonumber\\
&=&  A \; B + \sqrt{A^2-1} \;\sqrt{B^2-1}.
\end{eqnarray}
\begin{eqnarray}
\tanh\left(\tanh^{-1}A+\tanh^{-1}B\right) &=&  
{\tanh\left(\tanh^{-1}A\right)  +\tanh\left(\tanh^{-1}B\right)\over 1 + \tanh\left(\tanh^{-1}A\right) \tanh\left(\tanh^{-1}B\right)}
\nonumber\\
&=& {A+B\over 1 + AB}.
\end{eqnarray}

This now implies the more subtle result
\begin{eqnarray}
\sech\left(\sech^{-1}A+\sech^{-1}B\right) &=&  {1\over \cosh\left(\cosh^{-1}(1/A)+\cosh^{-1}(1/B)\right) }
\nonumber\\
&=& {1\over  A^{-1} B^{-1} +\sqrt{A^{-2}-1} \sqrt{B^{-2}-1} }
\nonumber\\
&=& {AB\over 1 + \sqrt{1-A^2} \sqrt{1-B^2}}.
\end{eqnarray}

\clearpage


\begin{thebibliography}{69}

\bibitem{Merzbacher}
E.~Merzbacher,
\emph{Quantum Mechanics}, (Wiley, New York, 1965).

\bibitem{Mathews}
P.~M.~Mathews and K.~Venkatesan,
\emph{A textbook of Quantum Mechanics}, 
(McGraw-Hill, New York, 1978).

\bibitem{Singh}
J.~Singh,
\emph{Quantum Mechanics: Fundamentals and applications to technology}, 
(Wiley, New York, 1997).

\bibitem{Sanchez-Soto}
L. L. Sanchez-Soto, J. F. Cari\~nena, A. G. Barriuso, J. J. Monzon,
``Vectorlike representation of one-dimensional scattering'',
European Journal of Physics {\bf26} (2005) 469,
[arXiv:quant-ph/0411081].

\bibitem{pedagogical}
P. Boonserm and M. Visser,
``One dimensional scattering problems: 
A pedagogical presentation of the relationship between reflection and transmission amplitudes'',
Thai Journal of Mathematics, 
Special Issue (Annual Meeting in Mathematics, 2010), pages 83--97.
Online ISSN 1686-0209.
{\sf www.math.science.cmu.ac.th/thaijournal}

\bibitem{Peres}
Asher Peres,
``Transfer matrices for one-dimensional potentials'',
J. Math. Phys. {\bf24} (1983) 1110--1119.

 \bibitem{Kowalski}
Jacek M. Kowalski and John L. Fry,
``Tunneling in one-dimensional ideal barriers'',
J. Math. Phys. {\bf28} (1987)   2407--2415.

\bibitem{Korasani}
S.~Korasani and A.~Adibi,
``Analytical solution of linear ordinary differential equations by a differential transfer matrix method'',
Electronic Journal of Differential Equations {\bf79} (2003) 1--18.

\bibitem{SS1}
A. G. Barriuso, J. J. Monzon, L. L. Sanchez-Soto, J. F. Cari\~nena,
``A vectorlike representation of multilayers'',
JOSA A {\bf21} (2004) 2386-2391,      
doi:10.1364/JOSAA.21.002386,
[arXiv:physics/0403140].

\bibitem{SS2} 
A. G. Barriuso, J. J. Monzon, L. L. Sanchez-Soto, J. F. Cari\~nena,
``Geometrical aspects of first-order optical systems'',
 Journal of Optics A: Pure and Applied Optics {\bf7} (2005) 451--456,
 doi:  10.1088/1464-4258/7/9/002,
 [arXiv:physics/0506112 [physics.optics]].
 


\bibitem{Bounds0}
M.~Visser,
  ``Some general bounds for 1-D scattering'',
  Phys.\ Rev.\  A {\bf59}  (1999) 427--438
  [arXiv: quant-ph/9901030].
  
  
\bibitem{Bounds-beta}
P. Boonserm and M. Visser,
``Bounding the Bogoliubov coefficients'',
Annals of Physics {\bf323} (2008) 27792798
[arXiv: quant-ph/0801.0610].
  
\bibitem{Bounds-greybody}
  P.~Boonserm and M.~Visser,
  ``Bounding the greybody factors for Schwarzschild black holes'',
  Phys.\ Rev.\  D {\bf 78} (2008) 101502
  [arXiv:0806.2209 [gr-qc]].
  
 \bibitem{Bounds-mg}
  P.~Boonserm and M.~Visser,
  ``Transmission probabilities and the Miller-Good transformation'',
  J.\ Phys.\ A  {\bf 42} (2009) 045301
  [arXiv:0808.2516 [math-ph]].

\bibitem{Bounds-analytic}
  P.~Boonserm and M.~Visser,
  ``Analytic bounds on transmission probabilities'',
  Annals of Physics {\bf 325} (2010) 1328--1339
[arXiv:0901.0944 [gr-qc]].
doi:~10.1016/j.aop.2010.02.005

\bibitem{Shabat-Zakharov}
  P.~Boonserm and M.~Visser,
  ``Reformulating the Schrodinger equation as a Shabat--Zakharov system'',
  Journal of Mathematical Physics {\bf51} (2010) 022105
  [arXiv:0910.2600 [math-ph]].

 \bibitem{Bounds-thesis}
 P.~Boonserm,
 \emph{Rigorous bounds on Transmission, Reflection, and Bogoliubov coefficients},
 (PhD thesis, Victoria University of Wellington, July 2009),   arXiv:0907.0045 [math-ph].
 
 \bibitem{QNF:analytic}
  P.~Boonserm, M.~Visser,
  ``Quasi-normal frequencies: Key analytic results'',\\
  Journal of High Energy Physics {\bf 1103} (2011) 073 
  [arXiv:1005.4483v2 [math-ph]].
  
\bibitem{Abeles1}
F. Abeles, ``La th\'eorie g\'en\'erale des couches minces'', Le Journal de Physique et le Radium, {\bf11}, 307--310 (1950).

\bibitem{Abeles2}
F. Abeles, 
``Recherches sur la propagation des ondes \'electromagn\'etiques sinusiodales dans les milieux stratifi\'es. 
Applicacion aux couches minces'', 
Annales de Physique {\bf5} (1950), 596--640, 706--782.

\bibitem{Lekner}
John Lekner,
\emph{Theory of reflection},
(Martinus Nijhoff, Dordrecht, 1987). 

\bibitem{Furman}
Sh. A. Furman and A.V. Tikhonravov, 
\emph{Basics of optics of multilayer systems},
(Edition Frontieres, Gif-sur-Yvette, 1992)
  
 
\end{thebibliography}
\end{document}